**A Visible Metamaterial with Low Loss Made by Bottom-Up Self-Assembly**

Boyi Gong, Xiaopeng Zhao*, Zhenzhen Pan, Sa Li, Xiaonong Wang, Yan Zhao, and Chunrong Luo

[*] Prof. X. P. Zhao, Dr. B. Y. Gong, M. Z. Z. Pan, M. S. Li, M. X. N. Wang, M. Y. Zhao, Prof. C. R. Luo
Smart Materials Laboratory, Department of Applied Physics,
Northwestern Polytechnical University, Xi' an 710129, P. R. China
E-mail: (xpzhao@nwpu.edu.cn)

Since the introduction of the artificially designed negative-index metamaterial (NIM) introduced in 2001,[1] its extraordinary electromagnetic properties, which cannot be attained from naturally occurring materials, have continuously attracted many researchers to study them.[2-4] Various types of NIMs with the resonant frequencies shifted from gigahertz all the way to higher frequencies have been actualized over the past decade. For the actual applications, the most fascinating and significant goal of research on NIMs is to achieve negative refraction at visible wavelengths. According to the effective medium theory, the interior structural unit of NIMs must be on a scale much smaller than the operating wavelength.[5] However, manufacturing NIMs with very short resonant wavelengths is difficult. Up to now, man-made NIMs have been developed at near-infrared and red-light wavelengths. In these wavelength ranges, the fishnet structure has been widely studied as a typical design since its first appearance in 2005.[6, 7] The negative refraction for fishnet NIM has been established at wavelengths of about 1.6 μm,[8] 1μm,[9] and recently close to 700-nm red-light wavelength.[10] Nearly all of the current NIMs have been fabricated with top-down etch technology, such as electron beam lithography and focused ion beam milling.[11] Duo to the intricate procedure, high cost and small size limit of etch fabrication, however, achieving smaller resonant structures by this method is impossible. Thus realization of NIMs at visible wavelengths remains a bottle-neck problem. In addition, the large inherent losses of NIMs seriously weaken their novel properties. A number of related studies have suggested that the most effective way to compensate for losses is to incorporate a gain material into the NIM structure. However, most studies to date have focused on the feasibility of gain via numerical simulations. Ref. [12] numerically investigated the addition of a gain medium into a fishnet NIM operating at 1.5 μm, whereas refs. [13] and [14] reported the response of a fishnet structure with gain at a working wavelength of 710 nm. However, the experiments on gain are scarce. A representative work in ref. [15] stated that an active medium, Rhodamine 800, can be embedded into a fishnet NIM operating at about 740-nm wavelength and then stimulated by a pulse laser, as a result substantially boosting transmission. As discussed above, the fabrication difficult and the large losses contribute to the extremely slow development of visible NIMs. Therefore a new fabrication method is necessary in attempting to replace the conventional etch technology.

Bottom-up electrochemical deposition has been developed for recent several years, and this method has been used to successfully fabricate metamaterials in some studies,[16–19] including our work where we successfully obtained near-infrared NIMs.[20–22] This fabrication method is favorable to produce NIMs with smaller size structural units that cannot be obtained from etch fabrication. Direct adoption of electrochemical deposition may result into disordered nanostructured units, if combing with a template-assisted self-assembled process, NIMs with periodic structures can be realized.[23] In this study, we fabricate a visible NIM with low-loss by template-assisted and self-assembled electrochemical deposition, producing a three-layer structure consisting of two mesh metal films asymmetrically on the opposite sides of a dielectric layer. The asymmetry property of the multilayer structure originates from the inherent characteristic of the bottom-up fabrication process. Its negative-index behavior at visible wavelengths can be validated both through numerical simulation and experimental data analysis. In order to overcome the large losses, an optically amplifying active medium,

Rhodamine B (Rh B), is incorporated in the dielectric layer. Control experiments show that the samples with Rh B have higher transmission magnitudes and better flat-lens focusing effects than samples without Rh B.

Through this approach, we produced an asymmetric silver (Ag)–polyvinyl alcohol (PVA)–Ag nanostructure on a glass substrate. Three issues must be addressed when applying this fabrication method: determination of whether or not the asymmetric structure can achieve negative-resonance behavior in theory, optimization of the fabrication process, and suppression of large inherent losses. Provided that the degree of asymmetry is kept within a moderate range, its resonance behavior can realize a negative index at visible wavelengths, as confirmed by after simulations according to the effective medium theory. Next, from our numerical analysis for compensating losses, we found that incorporating a gain medium into the structure can increase transmission, improve the performance of negative refraction, and exert nearly no effect on the negative-index frequency range. For comparison, control experiments have been done. Rh B was mixed into half of our samples during fabrication. The higher transmission and better flat-lens focusing effects verify the gain effect of Rh B. Hence, the results of this study have significant effects on visible NIMs. We have made a breakthrough in the fabrication of visible NIMs with smaller sizes by template-assisted and self-assembled electrochemical deposition. An asymmetric NIM was successfully fabricated, with negative-index wavelengths shifted to 650 and 550 nm. The inclusion of Rh B effectively compensated for losses and enabled better negative-index behaviors. Given its low cost, relatively simple fabrication process, and mass-production capability, this fabrication method offers new prospects for further developing visible NIMs.

Numerical simulations were performed using the commercial package COMSOL Multiphysics based on a finite integration technique. Silver does not behave as a perfect conductor at higher frequencies, and its dielectric function should be characterized by Drude model with a plasma frequency of $\omega_p = 1.37 \times 10^{16}$ $s^{-1}$ and a collision frequency of $\omega_c = 8.5 \times$ $10^{13}$ $s^{-1}$.[24] The standard algorithm of retrieving effective constitutive parameters for asymmetric structures was applied.[25] The refractive index (n) and impedance (z) were first retrieved from the four scattering coefficients (S11, S21, S22 and S12), and then permeability ($\mu$) and permittivity ($\varepsilon$) were obtained according to the equations of $n=\sqrt{\varepsilon\mu}$ and $z=\sqrt{\mu/\varepsilon}$. All of the geometrical dimensions of our model and the polarization configuration of the electromagnetic wave are shown in **Fig. 1**. The plane wave is at normal incidence with the electric field polarized in y-direction. The rectangular area surrounded by yellow curve is the structural unit. The holes in the silver films were designed to be polygonal, instead of rounded, because mutual extrusion of these dense holes distorts their rounded shape during fabrication. Silver films with a thickness (t) of 30 nm are separated by a 40-nm-thick PVA layer, the radius of the holes (r) is 70 nm and the central distance between the neighboring holes (w) is 235 nm. To describe the disalignment degree, the upper silver film is assumed to move simultaneously along the x-direction with Δx and along the y-direction with Δy (assuming Δx = Δy in our simulations) with respect to the lower silver film.

For convenience, we numerically analyze the electromagnetic response of the asymmetrical structural unit when increasing its disalignment degree. The retrieved effective μ, ε and n are plotted in **Figs. 2**(a), **2**(b), and **2**(c), respectively. The figure of merit (FOM), defined as - Re(n)/Im(n), which characterizes the inherent loss, is also shown in **Fig. 2**(d). It can be clearly seen in **Fig. 2**, when the disalignment degree Δx is kept at 20 nm the negative refractive index at visible wavelengths can be still maintained. However, we note here that the value of FOM decreases greatly with increasing Δx, which indicates that the inherent losses become more serious with the growth of Δx. This can be explained that the more and more portion of the holes is covered by the silver film on the other side of the PVA layer when increasing Δx, and the resonance between the two silver films becomes weak. We can conclude two important aspects from the curves in **Fig. 2**: (i) the asymmetrical structure can really achieve negative refraction at visible frequencies if Δx is controlled within a moderate range, and (ii) the large inherent losses seriously weaken the performance of the resonant structure. Hence, incorporating a gain medium into the structure is considered to compensate for losses. When the applied time-varying electric-field is expressed in term of exp(-jωt) time-dependence as $E = E(r)e^{-j\omega t}$, the imaginary part of permittivity is positive for the loss medium and negative for the gain medium. We incorporated the gain medium into a single unit cell with Δx=Δy=20 nm during simulation and analysis. The capability of gain is characterized by $\varepsilon''_{\mathrm{gain}}$, which is the imaginary part of permittivity for the gain medium. The effects of mixing gain on the electromagnetic response of the resonant structure are also simulated. The retrieved refractive index and the calculated FOM are shown in **Figs. 3**(a) and **3**(b), respectively. It can be found that the addition of a gain does not change the negative-index frequency range but suppresses the losses and allows better functionality. These numerical results demonstrate the feasibility of gain to compensating for losses and offer theoretical guidance for our subsequent experiments.

The actual samples were fabricated via template-assisted self-assembled electrochemical deposition, as illustrated in **Fig. 4**. In brief, the fabrication steps may be described as follows: (i) Nanometer-scale polystyrene (PS) spheres are first prepared and arranged periodically on a glass substrate with template-assisted method; [5] (ii) a layer of silver was electrochemically deposited into the apertures among these PS spheres; (iii) all the PS spheres are dissolved with chloroform and the remaining chloroform is dissolved with alcohol to form the lower layer of the silver film with holes on the glass substrate; (iv) the PVA solution is spin-coated onto the silver film and the sample is placed in a dry box for solidify; (v) the second layer of silver films with holes is formed on the solidified PVA layer by repeating the steps outlined from (ii) to (iv). A three-layer nanostructure based on the Ag-PVA-Ag sandwich is finally obtained.

Scanning electron microscopic images of two groups of samples with different sizes operating at the red- and green-light wavelengths are displayed in **Figs. 5**(a) and **5**(b), respectively. Rh B, an active medium, was incorporated into the PVA dielectric layer in order to compensate for losses, and its gain efficiency was investigated through control experiments. Each group includes two types of samples: those with and without Rh B. Transmission experiments were performed using a UV-4100 spectrophotometer. **Fig. 6**(a) plots the transmission spectra of one group of samples operating at red-light wavelengths. The black curve represents the sample without Rh B while the red curve corresponds to the sample with Rh B. Both transmission peaks are located at $\lambda$ = 650 nm but the peak value is boosted from 8.2% to 12.6% after inclusion of Rh B. **Fig. 6**(b) shows the transmission spectra of another group of samples with working frequences at green-light wavelengths, the black and green curves represent samples without and with Rh B, respectively. Both transmission peaks are maintained at about $\lambda$ = 550 nm, but the peak value increases from 14.1% to 26% when Rh B is incorporated into the samples. Later on, Flat-lens focusing experiments on these samples were also accomplished via our own design setup, as shown in **Fig. 5**(c). When a beam of polychromatic light is scattered from a xenon-lamp and passes through a special monochromator, it is transformed into monochromatic light with the same wavelength as the negative-index wavelength of the measured sample and then focused at a point by a convex-lens. The scattered light from the point source is focused again after passing through a negative-index flat sample, and the focused point can be detected with an optical-fiber spectrometer fixed on a micro-positioner. The phenomenon produced is called the flat-lens focusing effect, a unique property of NIMs that can be used to discriminate NIMs from traditional materials with a positive refractive index. Similar experiments can be found in previous literatures.[26,27] **Fig. 6**(a) and **Fig. 7**(a) employ the same sample as shown in **Fig. 5**(a), **Fig. 6**(b) and **Fig. 7**(b) use the same sample as displayed in **Fig. 5**(b). According to the measured results shown in **Fig. 7**, the negative-index properties of the transmission peaks plotted in **Fig. 6** can be unambiguously evaluated. For example, as depicted in **Fig. 7**(a), a flat-lens focusing phenomenon is observed with a source wavelength of 650 nm, it implies the negative-index property of the transmission peak at 650 nm wavelength as displayed in **Fig. 6**(a) and also proves the sample shown in **Fig. 5**(a) has a negative index around this wavelength. In **Fig. 7**, the horizontal axis represents the position of the measured output light beam after it passes through the flat sample and leaves the sample surface (set as the zero point). The vertical axis represents the normalized values of the measured intensity of the output light beam. **Fig. 7**(a) shows a comparison of the results for samples with Rh B (red curve) and without Rh B (black curve). The locations of both focus points are maintained at the same position but the focus intensity is more evident for the samples with Rh B. **Fig. 7**(b) displays a comparison of the results for samples with (green curve) and without (black curve) Rh B. The source wavelength is 550 nm, which corresponds to the transmission peak in **Fig. 6**(b). Likewise the locations of the two focus points are the same and the sample with Rh B exhibits better flat-lens focusing effects. The combined results in **Figs. 6** and **7** demonstrate the negative-index property of the asymmetric Ag-PVA-Ag nanostructure at visible frequencies. The low transmissions of the samples without Rh B indicate large losses, agreeing with the previous numerical analysis. The transmission increases, and the flat-lens focusing effect becomes more obvious when Rh B is added into the PVA layer. These results denote Rh B as a gain medium can compensate for losses and that we have fabricated a visible low-loss NIM.

As discussed above, we report an asymmetrical metal-dielectric-metal nanostructured metamaterial fabricated by template-assisted and self-assembled electrochemical deposition. The negative-index property of this structure in the visible range was numerically verified. Subsequent experiments, such as the transmission measurements and the flat-lens focusing effects, further demonstrated the negative refraction properties, and mixing an active medium Rh B into the samples availably compensates for the losses. In the experiments, we prepared red- and green-light-responsive samples. Each case included the samples with and without Rh

B as the control. In the case of red-light samples, the transmission peak at λ = 650 nm increases from 8.2% to 12.6% with the addition of Rh B. The intensity of flat-lens focusing similarly increases remarkably. For green-light samples, the transmission peak at λ = 550 nm shifts from 14.1% to 26%, and the intensity of flat-lens focusing also increases obviously after addition of Rh B. These results indicate that Rh B can be considered an effective gain for overcoming losses. In summery, the fabrication approach employed in this study successfully achieves two breakthroughs: fabrication of a visible NIM and suppression of inherent losses. The technique described here also allows large-scale, mass-production and low-cost manufacturing process.

**Acknowledgements**

This work was supported by the National Nature Science Foundation of China under Grant No. 50936002, 11174234, the National Key Scientific Program of China (under project No. 2012CB921503).

Received: ((will be filled in by the editorial staff))
Revised: ((will be filled in by the editorial staff))
Published online: ((will be filled in by the editorial staff))

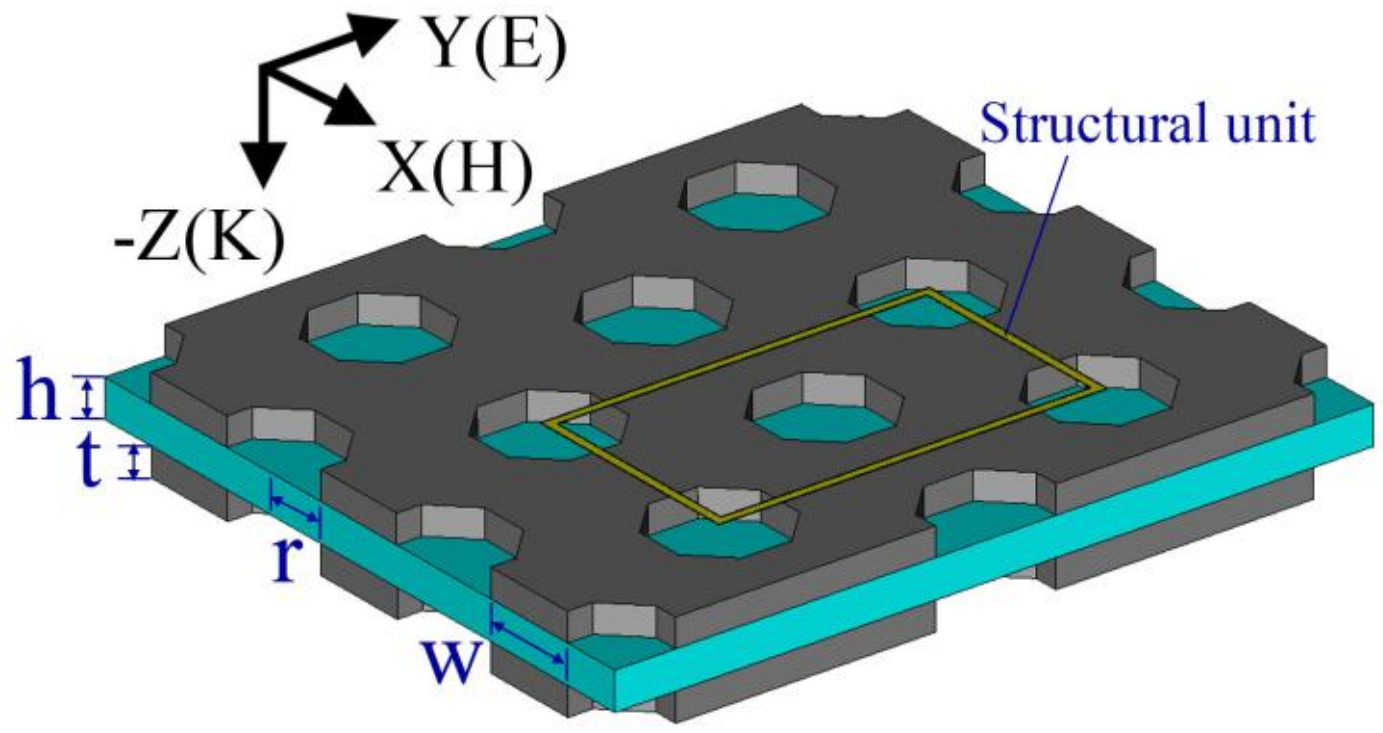

**Figure 1.** Schematic of an array of three-layer asymmetrical nanostructures and the polarization configuration of an electromagnetic wave. The geometrical parameters are w = 235 nm, r = 70 nm, t = 30 nm, and h = 40 nm, and the electromagnetic wave is at normal incidence with the electric field along the y-direction. The rectangular part surrounded by yellow curve is the structural unit.

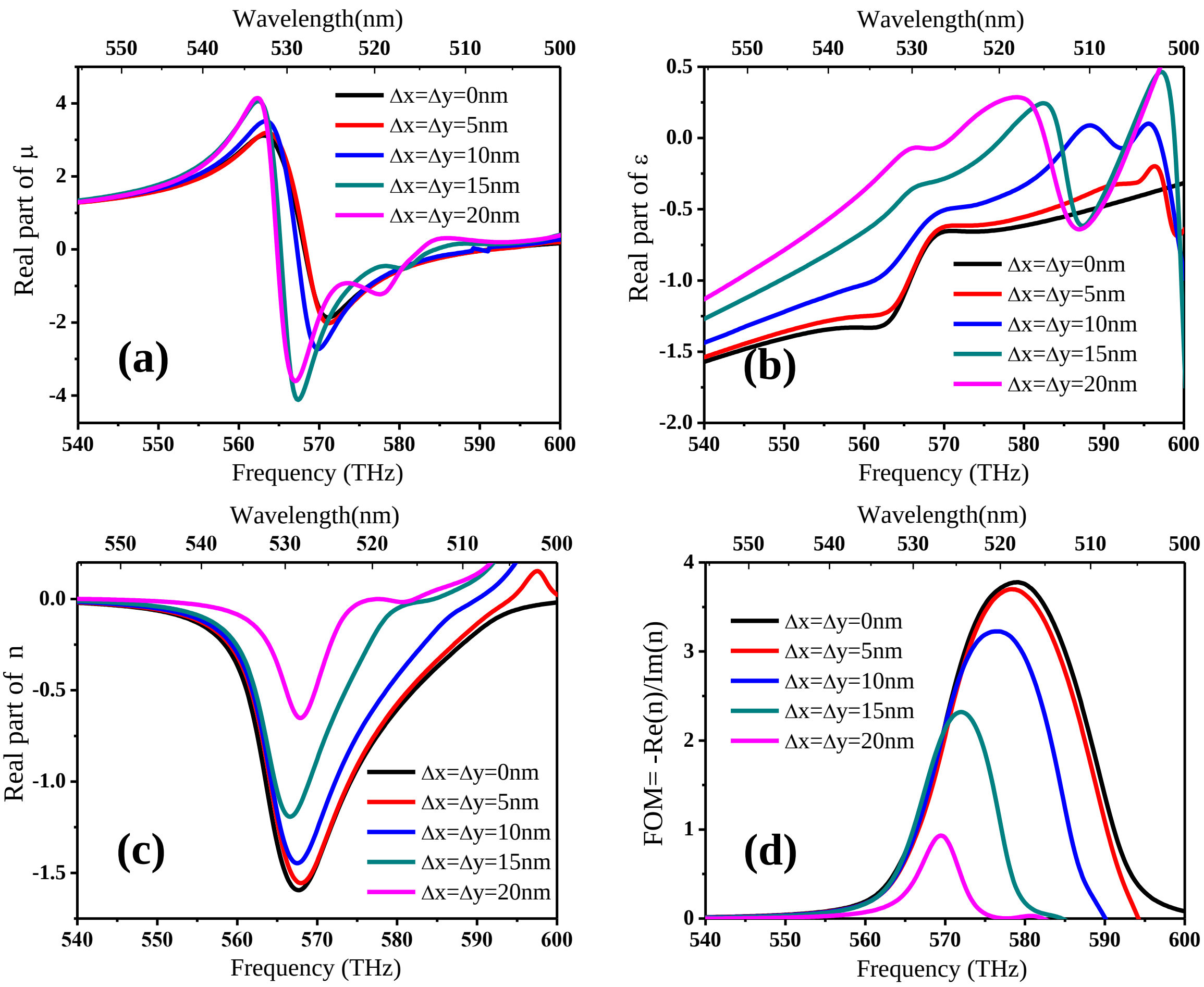

**Figure 2.** Effective constituent parameters with varying degrees of disalignment are retrieved from the simulated S parameters, such as the real parts of (a) permeability, (b) permittivity, (c) refractive index, and (d) the value of FOM, defined as -Re(n)/Im(n).

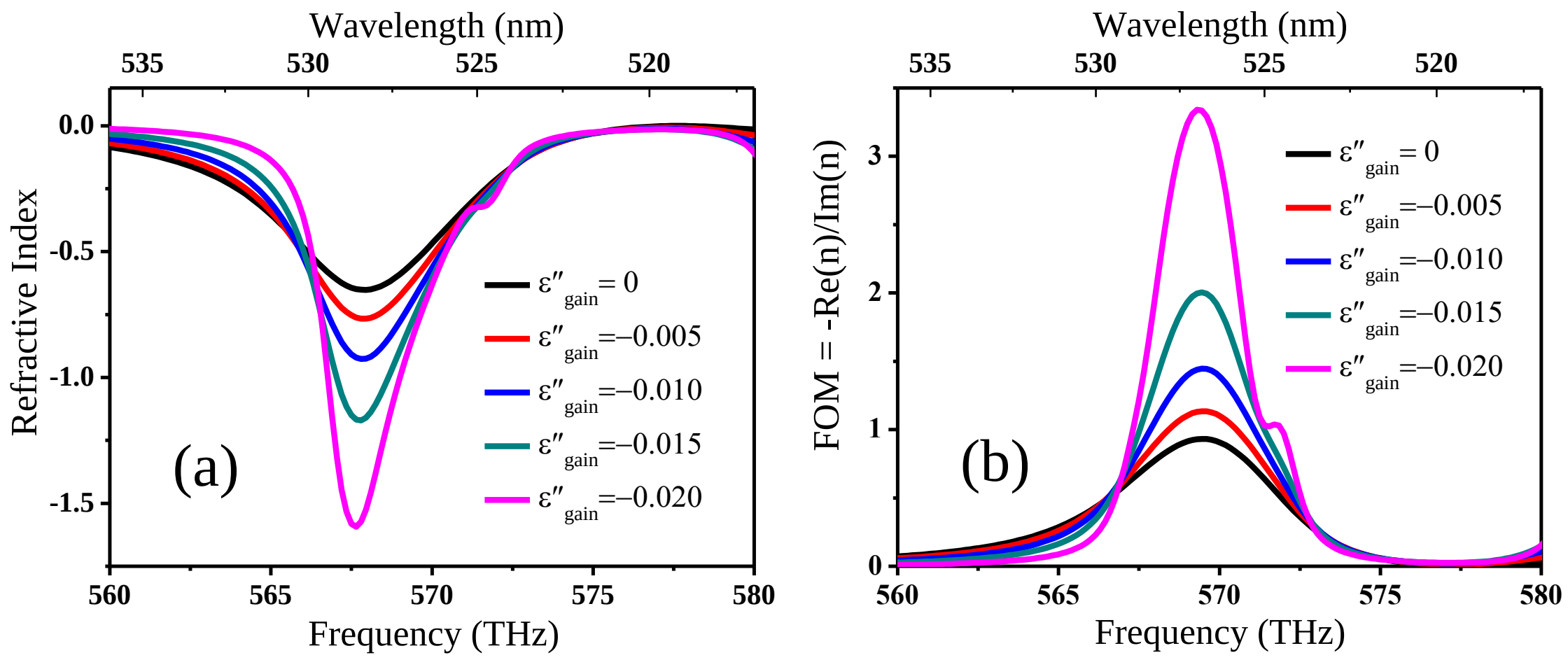

**Figure 3.** Response curves obtained after incorporating gain into a single unit cell with $\Delta x = \Delta y = 20$: (a) the retrieved refractive index, and (b) the calculated value of FOM.

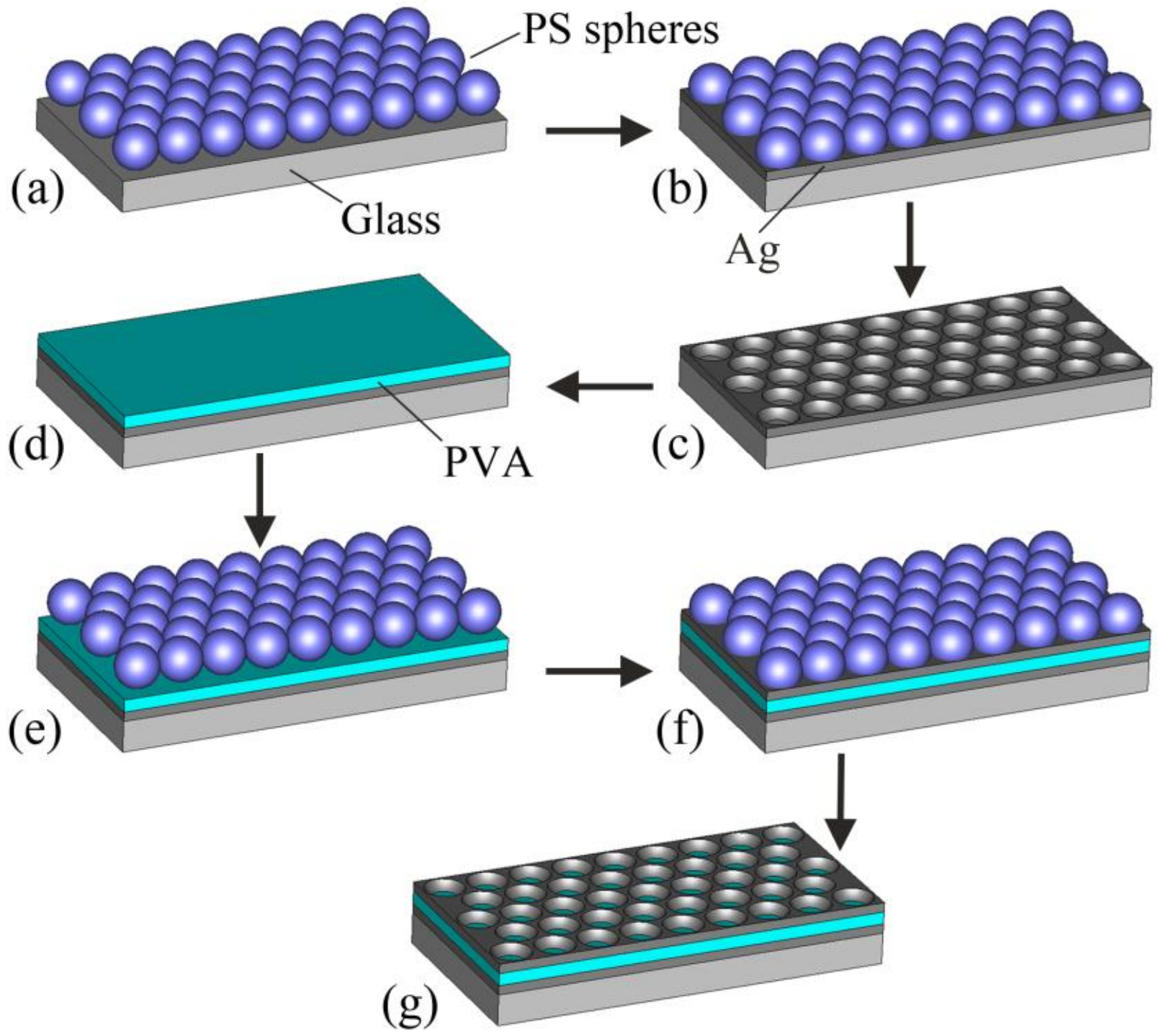

**Figure 4.** Fabrication steps. (a) Nanometer-scale PS spheres were periodically arranged on a glass substrate. (b) A silver layer was electrochemically deposited into the apertures among these PS spheres. (c) After removsal of the PS spheres with chloroform and alcohol, the lower layer of the silver film with holes was obtained. (d) PVA solution was spin-coated on the silver film and can be solidified in a dry box. (e) The second layer of PS spheres was periodically arranged on the PVA layer. (f) The upper layer of silver was also deposited. (g) In a manner similar to (c), the three-layer Ag-PVA-Ag nanostructure was finally formed by dissolving all of the PS spheres.

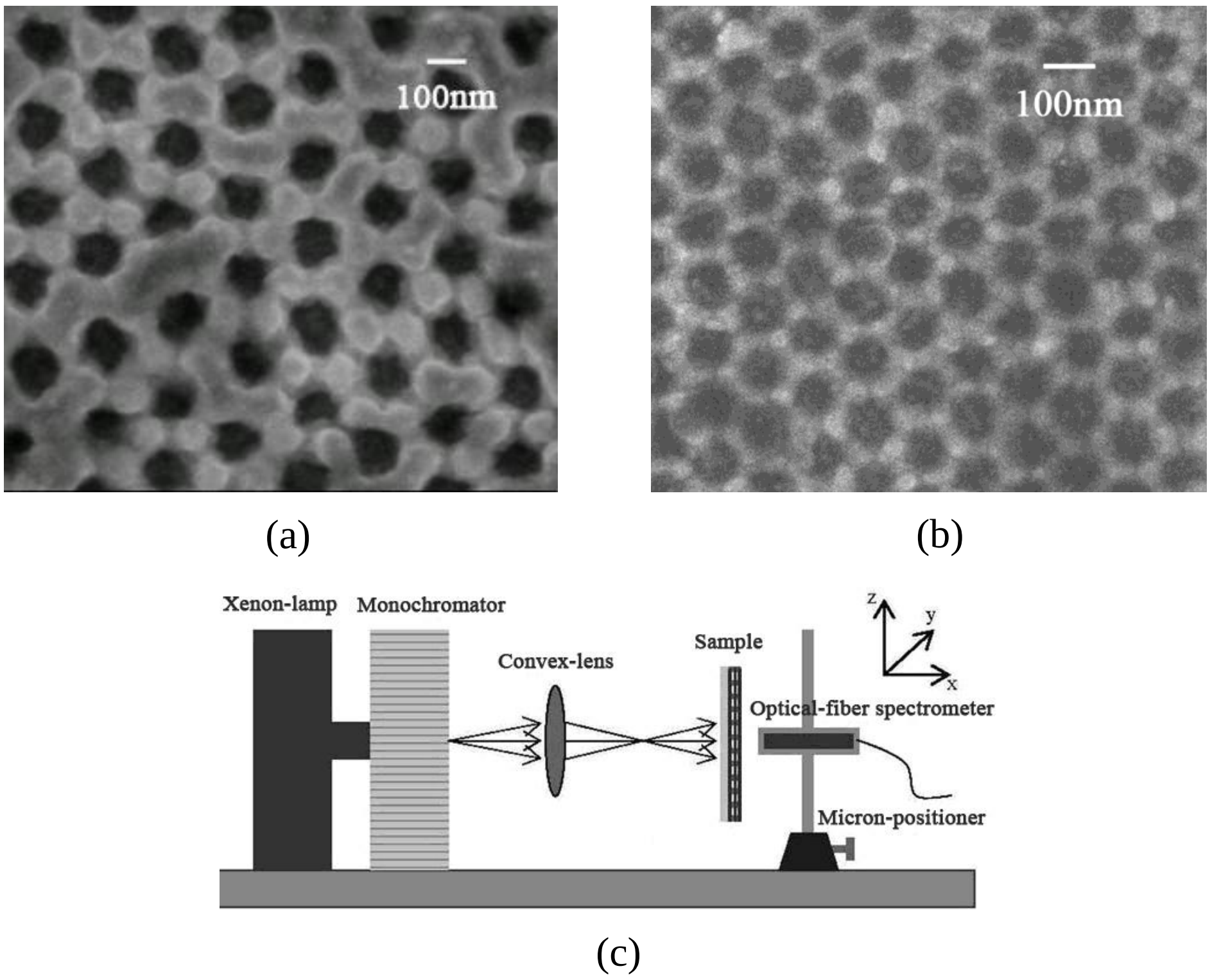

**Figure 5.** Top-view electron micrographs of the fabricated samples operating at red-light and green-light wavelengths are shown in (a) and (b), respectively. (c) Setup for the flat-lens focusing experiment.

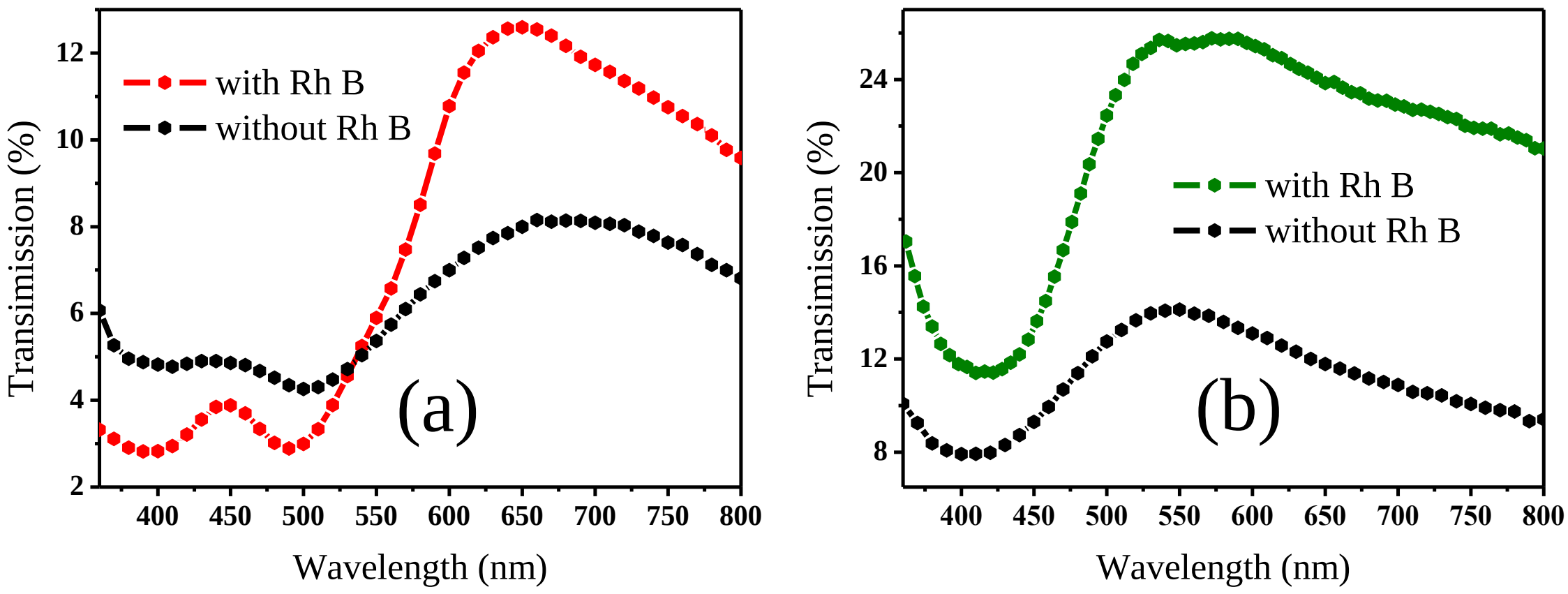

**Figure 6.** Measured transmission spectra of samples at (a) red-light and (b) green-light wavelengths. The black and green curves correspond to the samples without and with Rh B, respectively.

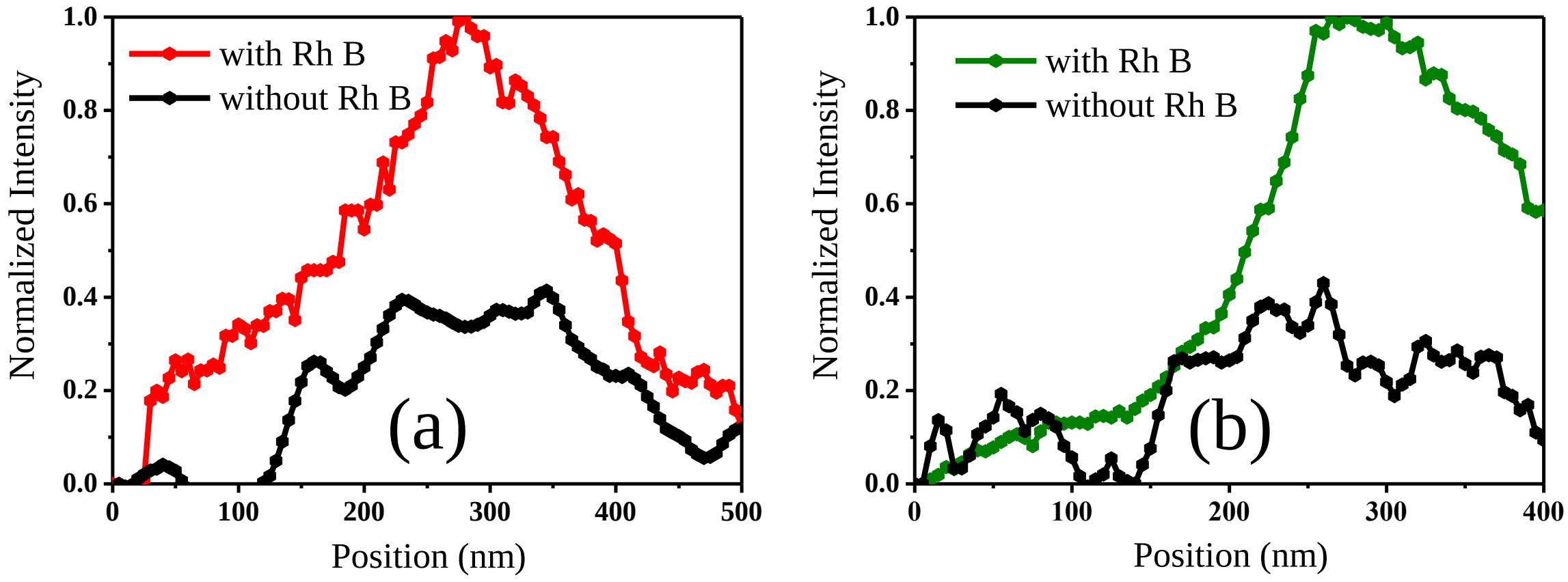

**Figure 7.** Measured results of flat-lens focusing experiments. The horizontal axis represents the positions of the measured output light beam when it passes through the flat sample and leaves the sample surface (zero point is set at the sample surface). The vertical axis represents the measured normalized intensity of the output light beam. (a) and (b) correspond to two different groups of samples operating at 650-nm and 550-nm wavelength, respectively. The black curves denote the results for samples without Rh B while the red and green curves indicate samples with Rh B.